\documentclass[showpacs,preprintnumbers,amsmath,amssymb,floatfix,elsart,prd]{revtex4}

\usepackage{graphicx}
\usepackage{dcolumn}
\usepackage{bm}

\newcommand{\beq}{\begin{equation}}
\newcommand{\eeq}{\end{equation}}
\newcommand{\bey}{\begin{eqnarray}}
\newcommand{\eey}{\end{eqnarray}}

\begin{document}

\title{Strong Lensing of a  Regular Black Hole with an Electrodynamics Source}

\author{Tuhina Manna}
\email{ tuhinamanna03@gmail.com} \affiliation{Department of Mathematics(Commerce Eve)
              St.Xavier's College, 30 Mother Teresa Sarani, Kolkata-700016,
India}

\author{Farook Rahaman}
\email{rahaman@associates.iucaa.in} \affiliation{Department of
Mathematics, Jadavpur University, Kolkata 700032, West Bengal,
India}

\author{Jhumpa Bhadra}
\email{bhadra.jhumpa@gmail.com} \affiliation{  Department of
Mathematics,
              Heritage Institute of Technology, Kolkata - 700107,
              India}
\author{ Sabiruddin Molla}
\email{sabiruddinmolla111@gmail.com} \affiliation{Department of
Mathematics, Jadavpur University, Kolkata 700032, West Bengal,
India}\author{   Hasrat Hussain Shah }
\email{hasrat@mail.ustc.edu.cn} \affiliation{CAS Key Laboratory for Research in Galaxies and Cosmology, Department of Astronomy,
University of Science and Technology of China, Hefei 230026, China and \\
Department of Mathematical Sciences, Baluchistan University of Information Technology, Engineering and Management Sciences, Quetta 87300, Pakistan.}

\date{\today}

\begin{abstract}
In this paper we have investigated the gravitational lensing phenomenon in the strong field regime for a regular, charged, static     black holes with non-linear  electrodynamics source. We have obtained the angle of deflection and compared it to a  Schwarzschild black hole and Reissner Nordstr\"{o}m black hole with similar properties. We have also done a graphical study of the relativistic image positions and magnifications.  We hope that this method may be useful in the detection of non-luminous bodies like this current black hole.
\end{abstract}


\maketitle

\section{Introduction}

 Gravitational Lensing is one of the fundamental observational proofs of Einstein's General Theory of Relativity. The idea that even light can be deflected by the influence of gravity dates back to the era of Newtonian Theory of gravity as a 'gravitational force'. This phenomenon has been suspected by Sir Issac Newton, Laplace and Soldner, to mention a few. However they based their calculations on Newtonian gravity, unlike Einstein who calculated the deflection of light by the Sun using his own theory, as being of the magnitude of 1.7 arcsecs . Later this was confirmed by Lord Eddington's expedition of 1920. Zwicky and Schmidt predicted the deflection of light can be used to study galaxies and quasars \cite{NAR}.    The lensing effect is  now  widely used for investigating non-luminous objects such as extra solar planets, distant stars, wormholes \cite{FR}, in detection of dark matter, estimation of cosmological parameter, detection of gravitational waves, and cosmic censorship conjuncture . The gravitational lensing theory begun with development in the
weak field approximation, which has been successfully employed to explain astronomical observations \cite{Schneider}. Although, weak approximation becomes invalid in close proximity of a compact star, around which the light can orbit one or several times before emerging. However, in the last few decades, the gravitational lensing in strong field regime including lensing properties near photon sphere, has gained much attention. It has   proven  to be  a  useful tool in detecting non luminous objects such as extra solar planets, dark matter, estimation of cosmological
parameter, gravitational waves and black holes since the observed
effects of gravitational lensing like multiple images,  arcs,
rings, or any distortion of images are quite easily detectable by
using high resolution imaging by very long baseline interferometry
(VLBI).  Darwin \cite{Darwin} was the pioneer of the work in
strong gravitational lensing by compact objects with a photon
sphere, like black holes   and naked singularities. It has been
extensively investigated which indicates  the presence of
supermassive black hole  at the center of galaxies.  Strong lensing is
related to photons passing close to the photon sphere and being
able to circumvent multiple times the black hole before reaching an
observer. Virbhadra and Ellis \cite{Vibhadra} recently
investigated gravitational lensing in the strong field and
obtained the lens equation. They established that a source behind
a Schwarzschild black hole produce two infinite sets of
relativistic images on each side of the black hole. Subsequently, Bozza et
al. \cite{Bozza1,Bozza2} provided the formulation of the
gravitational lensing theory in the strong field limit for a
Schwarzschild black hole as a counterpart to the weak field
approach. Their method is universal and can be applied to any
asymptotically flat spacetime. They proved that the deflection
angle always diverges logarithmically when the minimum impact
parameter is reached. Afterwards Frittelli, Kling \& Newman
\cite{Frittelli} found an exact lens equation, giving integral
expressions for its solutions.  The same technique
is applied by Eiroa, Romero and Torres \cite{Eiroa} to a Reissner -  Nordstr\"{o}m black hole.  Virbhadra \& Ellis \cite{Ellis}, in a work, differentiated the main features of gravitational lensing by normal black holes and by naked singularities, analyzing the Janis, Newman, Winicour metric. They have mentioned the importance of their studies after providing a test for the cosmic censorship hypothesis.\par
In our study  we have considered a regular black hole studied by
Balart and Vagenas \cite{BH} having a non-linear electrodynamics
source with the distribution function as
$\sigma(r)=\exp\left(\frac{-q^2}{2Mr}\right)$.
  Regular black holes were first introduced by Bardeen in 1968 \cite{Bardeen} and since then they have been frequently discussed in the study of non linear electrodynamics \cite{AB}. Bardeen black hole is considered to be a magnetic solution of the Einstein equations coupled to non linear electrodynamics. In their 2011 paper Eiroa and Sanders \cite{Eiroaregular} have discussed the gravitational lensing of a regular Bardeen black hole, in terms of the self-gravitating monopole charge $g$ in non linear electrodynamics, using strong field limit, including a detailed analysis of the image positions and magnifications. They have shown that as $|g|/2M$ increases the relativistic images are formed closer to the black hole and also that the angular separation between the first image and the limiting value of the succession of images increases with increase in $|g|/2M$.   \par

  Badía et. al. \cite{horndeski} have investigated the gravitational lensing by non-rotating and asymptotically flat black holes in Horndeski theory. Zhang et. al. \cite{Zhang} have discussed black hole gravitational lensing for a spherically symmetric spacetime with torsion, in the generalized Einstein-Cartan-Kibble-Sciama theory of gravity. They have investigated different scenarios like the gravitational lensing in presence of horizon and photon sphere, in the absence of photon sphere but presence of horizon and absence of both. They have shown that in the absence of photon sphere and horizon the tends to be a constant $-\pi$ and becomes independent of torsion parameters. In recent paper by Chakraborty et. al. \cite{Chak} strong lensing phenomenon in three distinct black hole spacetimes were investigated namely the four dimensional black hole in presence of Kalb-Ramond ﬁeld, brane world black holes with Kalb-Ramond ﬁeld and black hole solution in f(T) gravity. Gravitational lensing by a negative Arnowitt-Deser-Misner (ADM) mass was studied in \cite{ADM}.  Asada et. al. worked on the gravitational lensing of exotic objects like Ellis wormholes recently and expressed the spacetime metric as a function of the inverse powers of the distance \cite{Asada}. Gravitational lensing is used as a tool to detect wormholes in outer regions of galactic halos in Bhar et.al. \cite{manna} using conformal motion. Several other papers \cite{others} remarkable study on the deflection angle by the Ellis wormhole have been studied in the strong field limit \cite{ellis}. Interesting phenomenon of retrolensing of charged black holes was studied  by Tsukamoto et. al.\cite{Tsukamoto} including some recent work on lensing of wormholes  by light passing through the wormhole throat \cite{Tsukamoto2} by the same author.  Also worth mentioning is a study on the gravitational lensing of a class of zero Ricci scalar wormholes by Rajibul et. al. \cite{Rajibul}.

We have done a
detailed analysis of the strong lensing phenomenon and compared
the angle of deflection as obtained with the angle of deflection
corresponding to the Schwarzschild black hole  and the Reissner Nordstr\"{o}m black hole.

The current paper
is divided in the following way: in section 2 we have given the outline of the
black hole solution under study, in section 3 we have done a
detailed analysis of the strong lensing and the compared the angle
of deflection with that of a similar Schwarzschild black hole  and Reissner Nordstr\"{o}m black hole. Next in section 3, the Image analysis we have discussed the image positions, critical curves and magnification.
Finally in section 4 we have discussed some concluding remarks. Throughout the paper we have used geometrized units taking $M=GM/c^2$

\section{Black hole   solutions with a nonlinear
electrodynamics source}
\label{sec:1}
 The pioneer numerous nonlinear electrodynamics model is known as Born-Infeld theory which described a finite electron self energy \cite{Born}. In quantum electrodynamics the nonlinear source can be produced due to virtual charged particle. The nonlinearity effect in the electrodynamics plays an important role only in strong electromagnetic field. Bardeen rooted the
the study with static spherically symmetric regular black hole solution \cite{Bardeen}. After his numerous study, many models have been proposed in order to discuss the regular black hole solution \cite{Borde,Barra,Bogo,Cabo,Hawy}.  Charged regular black holes are solutions of Einstein
equations having horizons and, their metrics together with their
curvature invariants are regular everywhere. They are dissimilar
to Reissner - Nordstr\"{o}m black holes which have
singularities at the origin. Few regular black holes violates the
strong energy condition somewhere in the spacetime where as some
of these satisfy the weak energy condition (WEC) everywhere.
Recently, L. Balart and E. C. Vagenas \cite{BH} constructed a nonsingular black hole solutions by coupling gravity to nonlinear electrodynamics source. We are following their consideration of the spacetime geometry.  \par
 In this study, we consider a spherically symmetric regular black hole spacetime in
general relativity coupled to a nonlinear electrodynamics. The line element describing the spherical symmetric charged regular black hole given by

\begin{equation}\label{eq:BH}
ds^{2} = -f(r)dt^{2}+\frac{1}{f(r)}dr^{2}+r^{2}(d\theta^{2}+\sin^{2}\theta d\phi)
\end{equation}
where following \cite{BH} we have taken the function $f(r)$ as,  \\
\begin{equation}
f(r)=1-\frac{2M}{r} \exp\left(\frac{-q^2}{2Mr}\right)
\end{equation}
and $q$ denotes the electric charge.\\
 It is easy to  check that  for   the given solution, WEC is satisfied. Therefore, the  energy-momentum is physically valid.

Here we have investigated the gravitational lensing of a black
hole with its source as non-linear electrodynamics model. The
horizons $r_\pm$ are obtained by solving for the real roots of
$f(r_\pm)=0$. They are
\begin{align}
r_+&=-\frac{q^2}{2M W \left(0,\frac{-q^2}{4M^2}\right)}\nonumber,\\
r_-&=-\frac{q^2}{2M W \left(-1,\frac{-q^2}{4M^2}\right)}.
\end{align}
where  W  denotes Lambert's W function.   The
corresponding electric field is given by

\begin{align}
E(r)=\frac{q}{r^2}\left(1-\frac{q^2}{8Mr}\right)\exp\left(\frac{-q^2}{2Mr}\right)
\end{align}

which is regular everywhere and asymptotically tends to
$E(r)=q/r^2+O(1/r^3) $ and represents extremal regular black hole
when $|q| = 1.213M$  \cite{BH}. We now investigate in the
next section the deflection angle of a path of light approaching
the black hole as a function of its distance of closest approach.

\section{Gravitational Lensing}
In this section we proceed to investigate the deflection angle of the black hole.
Comparing eq.~(\ref{eq:BH})  with the following metric
\begin{equation}
ds^{2}=-A(x)dt^{2}+B(x)dx^2+C(x)(d\theta^{2}+\sin^{2}\theta d\phi^{2})
\end{equation}

we get
\begin{flalign}
A(x)=&~1-\frac{2M}{x} \exp\left(\frac{-q^2}{2Mx}\right)\\
B(x)=&~\frac{1}{1-\frac{2M}{x} \exp\left(\frac{-q^2}{2Mx}\right)}\\
C(x)=&~x^2
\end{flalign}
 In order to start our study of the deflection angle, we first find a radial distance corresponding to which the angle of deflection is infinite. This is the  radius of photon sphere($x_m$) which is given as the largest positive root of the equation (see \cite{Ellis,V2}),
\begin{equation}
\frac{C'(x)}{C(x)}=\frac{A'(x)}{A(x)},
\end{equation}
i.e., it is given by the roots of the equation
\begin{equation*}
\frac{2x}{x^2}=\frac{\left(\frac{2M}{x^2}-\frac{q^2}{x^3}\right)\exp\left(-\frac{q^2}{2Mx}\right)}{1-\frac{2M}{x}\exp\left(\frac{-q^2}{2Mx}\right)},
\end{equation*}

which implies,
\begin{equation}
2x^2 \exp\left(\frac{q^2}{2Mx}\right)-6Mx+q^2=0.
\end{equation}

\begin{figure}\label{fig1}
\includegraphics[width=8 cm]{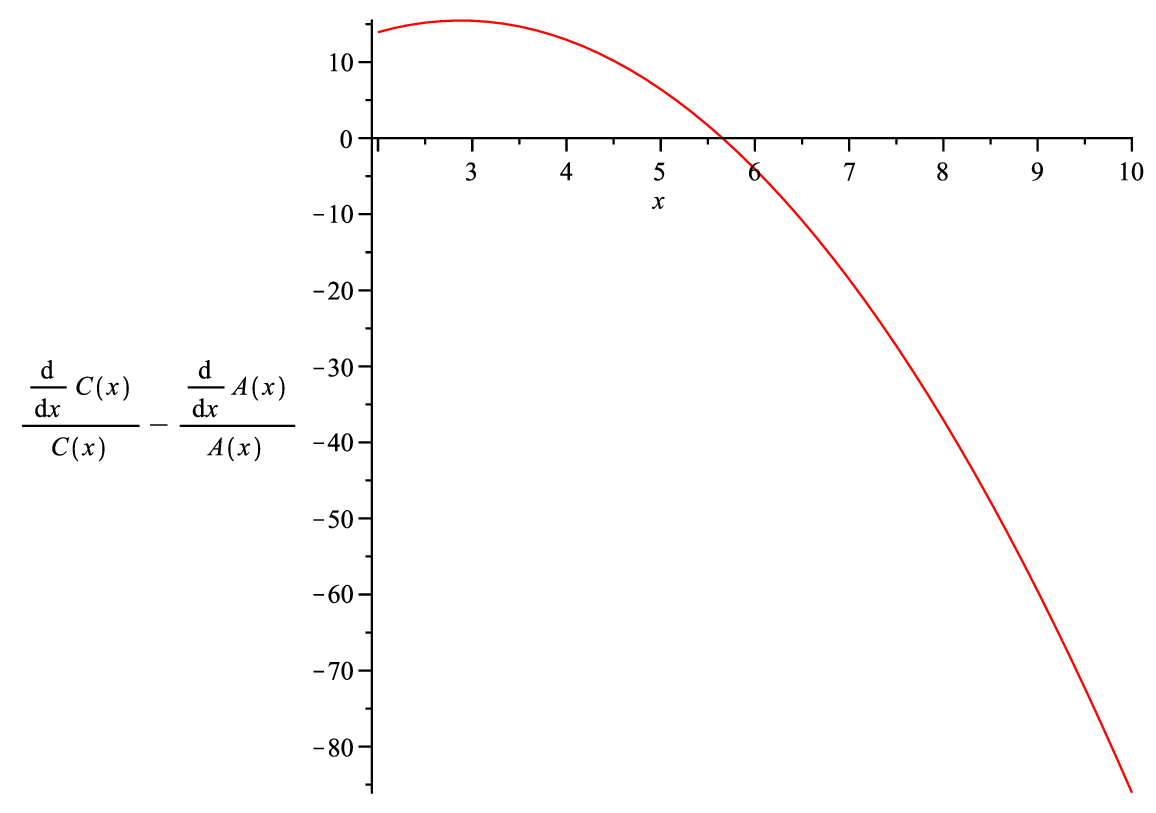}
\caption{The photon sphere($x_m$) corresponds to the largest root of the equation $\frac{C'(x)}{C(x)}=\frac{A'(x)}{A(x)}$. Hence, it is found graphically as the point where the function $\frac{C'(x)}{C(x)}-\frac{A'(x)}{A(x)}$ cuts the x-axis. The values of the parameters are $q=1 M_\odot $ and $M=2.5M_\odot$. }
\end{figure}

  Hence, it is found
graphically in the Fig 1 as the point where the function
$\frac{C'(x)}{C(x)}-\frac{A'(x)}{A(x)}$ cuts the x-axis. Using
$M=2.5 M_\odot$ and $q=1 M_\odot$ we have obtained graphically the value of
photon sphere as $x_m = 5.6559907$.

 Consider  the strong lensing method, where a photon from
some distant galaxy passes close to the event horizon of the black
hole with some impact parameter $'u'$.  General theory of
relativity claims that the immense gravity of  a black hole, causes the photon to move in a null geodesic, which would otherwise be a straight line path, as in the case of a flat spacetime in the absence of the black hole. Thus this phenomenon acts as a direct evidence of General theory of relativity and also as an observational tool to detect exotic objects.  Here we assume that the observer in located in an asymptotically flat spacetime. Many spectacular multiple images of a single source may be formed. When the gravitational lens is perfectly aligned with the observer, a ring-shaped image of the source is formed around actual position of the gravitational lens.\cite{Schneider}  There is
also a possibility that the photon may loop around the black hole several
times before spiralling out towards the observer, in case the
distance of closest approach $x_0$ is so less that it suffers a
deflection greater than $2\pi$, or it may loop around and fall in
the event horizon, in the extreme case when the distance of
closest approach decreases to equate the photon sphere.\\ Hence, proceeding in this direction we can get
by conservation of the angular momentum, the closest approach
distance is related to the impact parameter $u$
\begin{equation}\label{eq:u}
u=\frac{x_0}{\sqrt{1-\frac{2M}{x}\exp\left(\frac{-q^2}{2Mx}\right)}}.
\end{equation}
From the geodesic equation we can find the angle of deflection as (see \cite{W});
\begin{equation}\label{eq:alpha}
\alpha(x_0)=I(x_0)-\pi ,
\end{equation}
\begin{equation}\label{eq:I}
I(x_0)=\int_0^\infty\left(\frac{2\sqrt{B}dx}{\sqrt{C}\sqrt{\frac{CA_0}{C_0A}-\frac{C}{C_0}}}\right).
\end{equation}
Following \cite{Bozza} we take the following transformation of variables.
Let,
\begin{align}
y=&A(x)\nonumber\\
&=1-\frac{2M}{x}\exp\left(\frac{-q^2}{2Mx}\right),\\
y_0=&y(x~=~x_0)\nonumber\\
&=1-\frac{2M}{x_0}\exp\left(\frac{-q^2}{2Mx_0}\right),
\end{align}
and
\begin{align}
z=&\frac{y-y_0}{1-y_0}\nonumber\\
&=1-\frac{x_0}{x}\exp\left[\frac{q^2}{2M}\left(\frac{1}{x_0}-\frac{1}{x}\right)\right].
\end{align}
Thus, writing,
\begin{equation}
W=Lambert~W\left(0,\frac{-q^2(1-z)\exp(-\frac{q^2}{2Mx_0})}{2Mx_0}\right),
\end{equation}
we obtain
\begin{equation}
x=\frac{-q^2}{2MW}.
\end{equation}

Now, we can write the metric coefficients in terms of $z$ as
\begin{align}
A(z,x_0)=&1-\frac{2M(1-z)}{x_0}\exp\left(\frac{-q^2}{2Mx_0}\right),\\
B(z,x_0)=&\left[1-\frac{2M(1-z)}{x_0}\exp\left(\frac{-q^2}{2Mx_0}\right)\right],\\
C(z,x_0)=&\left[\frac{-q^2}{2MW}\right]^{2}.
\end{align}
Hence, the angle of deflection can be obtained as
\begin{equation}\label{eq:I(x0)}
I(x_0)=\int^1_0~R(z,x_0)f(z,x_0)dz,
\end{equation}
 where
\begin{equation}
R(z,x_0)=\frac{2\sqrt{A(x)B(x)}}{C(x)A'(x)}(1-A_0)\sqrt{C_0},
\end{equation}
and
\begin{equation}
f(z,x_0)=\frac{1}{\sqrt{y_0-[(1-y_0)z+y_0]\frac{C_0}{C}}}
\end{equation}

After  some  simplification we get
\begin{equation}\label{eq:Rzx)}
R(z,x_0)=\frac{2}{(1+W)}\exp\left({-\frac{q^2}{2Mx_0}-W}\right),
\end{equation}
and
\begin{equation}\label{eq:fzx}
f(z,x_0)=\left(1-\frac{4M^2x_0^2W^2}{q^4}-\frac{2M}{x_0}\exp\left(\frac{-q^2}{2Mx_0}\right)\left[1+\frac{4M^2x_0^2W^2(z-1)}{q^4}\right]\right)^{-1/2}.
\end{equation}
Now, putting $z=0$ we have,
\begin{equation}\label{eq:R0x}
R(0,x_0)=\frac{2\exp\left({-\frac{q^2}{2Mx_0}-W_0}\right)}{1+W_0},
\end{equation}
where
\begin{equation}
W_0=Lambert~W\left(0,\frac{-q^2\exp(-\frac{q^2}{2Mx_0})}{2Mx_0}\right).
\end{equation}

 Obviously, we can see from the expression of $R(z,x_0)$ in
eq. (\ref{eq:Rzx)}),  that it is regular in z, but $f(z,x_0)$ is
divergent at z=0, as seen in eq. (\ref{eq:fzx}). Following
\cite{Bozza} we can write the Taylor series expansion of
$f(z,x_0)$ up to second order and rename it as follows :
\begin{equation}\label{eq:f0}
f_0(z,x_0)=\frac{1}{\sqrt{p(x_0)z+q(x_0)z^2}}.
\end{equation}
where
\begin{flalign}\label{eq:pq}
p(x_0)=&\frac{(1-A_0)(C_0'A_0-C_0A_0')}{C_0A_0'},\nonumber\\
q(x_0)=&\frac{(1-A_0)^2}{2C_0^2A_0'^3}\left(2C_0C_0'A_0'^2+(C_0C_0''-2C_0'^2)A_0A_0'-C_0C_0'A_0A_0''\right).
\end{flalign}
Note that $p(x_0)=0$ at $x_0=x_m$, thus as the distance of closet approach reaches  the radius of photon sphere the integral $I(x_0)$ diverges.\\
 Thus,  simplifying eq. (\ref{eq:pq}) we get,
\begin{align}
p(x_0)=&\frac{2M}{2Mx_0-q^2}\left[2x_0-\left(6M-\frac{q^2}{x_0}\right)\exp\left(\frac{-q^2}{2Mx_0}\right)\right],\nonumber\\
\text{and}\nonumber\\
q(x_0)=&\frac{2M}{(2Mx_0-q^2)^3}\left[-4M^2x_0^3-2q^2x_0^2M+q^4x_0\right.\nonumber\\
&+\left.\left(2q^4M-12M^2q^2x_0+24M^3x_0^2\right)\exp\left(\frac{-q^2}{2Mx_0}\right)\right].
\end{align}

Hence, eq. (\ref{eq:f0})  becomes,

\begin{align}\label{f0zx0}
f_0(z,x_0)=&\left(\frac{2Mz}{2Mx_0-q^2}\left[2x_0-\left(6M-\frac{q^2}{x_0}\right)\exp\left(\frac{-q^2}{2Mx_0}\right)\right]\right.\nonumber\\
&\left.\frac{2Mz^2}{(2Mx_0-q^2)^3}\left[-4M^2x_0^3-2q^2x_0^2M+q^4x_0\right.\right.\nonumber\\
&\left.\left.+\left(2q^4M-12M^2q^2x_0+24M^3x_0^2\right)\exp\left(\frac{-q^2}{2Mx_0}\right)\right]\right)^{-1/2}.
\end{align}

In order to solve the integration we split (\ref{eq:I(x0)}) into two parts as :
\begin{equation}\label{eq:I}
I(x_0)=I_D(x_0)+I_R(x_0),
\end{equation}
where the divergent part is given as,
\begin{equation}\label{eq:Idiv}
I_D(x_0)=\int^1_0R(0,x_m)f_0(z,x_0)dz,
\end{equation}
and the regular part is given by,
\begin{equation}\label{eq:Ireg}
I_R(x_0)=\int_0^1g(z,x_0)dz,
\end{equation}
where
\begin{equation}\label{eq:g}
g(z,x_0)=R(z,x_0)f(z,x_0)-R(0,x_m)f_0(z,x_0).
\end{equation}
Simplifying   further  eqs. (\ref{eq:Rzx)}) , (\ref{eq:fzx}) , (\ref{eq:R0x}) , (\ref{f0zx0}) and (\ref{eq:g}) we get,
\begin{align}\label{eq:gzx0}
g(z,x_0)=&\frac{2\exp\left({-\frac{q^2}{2Mx_0}-W}\right)}{(1+W)}\left(1-\frac{4M^2x_0^2W^2}{q^4}-\frac{2M}{x_0}\exp\left(\frac{-q^2}{2Mx_0}\right)\left[1+\frac{4M^2x_0^2W^2(z-1)}{q^4}\right]\right)^{-1/2}\nonumber\\
&-\frac{2\exp\left({-\frac{q^2}{2Mx_0}-W_0}\right)}{1+W_0}\left(\frac{2Mz}{2Mx_0-q^2}\left[2x_0-\left(6M-\frac{q^2}{x_0}\right)\exp\left(\frac{-q^2}{2Mx_0}\right)\right]\right.\nonumber\\
&\left.+\frac{2Mz^2}{(2Mx_0-q^2)^3}\left[-4M^2x_0^3-2q^2x_0^2M+q^4x_0+\left(2q^4M-12M^2q^2x_0+24M^3x_0^2\right)\exp\left(\frac{-q^2}{2Mx_0}\right)\right]\right)^{-\frac{1}{2}}.
\end{align}

Considering terms that are first order in ($x_m-x_0$) we can write
\begin{equation}\label{eq:IDx0}
I_D(x_0)=-a\log\left(\frac{x_0}{x_m}-1\right)+b_D+O(x_m-x_0),
\end{equation}
where
\begin{align}\label{eq:abDqm}
a=&\frac{R(0,x_m)}{\sqrt{q_m}}\nonumber\\
b_D=&\frac{R(0,x_m)}{\sqrt{q_m}}\log\left(\frac{2(1-y_m)}{A_m'x_m}\right)\nonumber\\
\text{and} \nonumber\\
q_m~=&~q(x_0=x_m)=\frac{C_m(1-y_m)^2}{2y_m^2C_m'^2}\left(C_m''y_m-C_mA''_m\right).
\end{align}

Hence, we find from eq. (\ref{eq:abDqm})
\begin{align}
q_m=\frac{\frac{4M^2}{x_m^2}\exp\left(\frac{-q^2}{Mx_m}\right)}{8\left(1-\frac{2M}{x_m}\exp\left(\frac{-q^2}{2Mx_m}\right)\right)^2}\left[2x_m+\left(-4M+\frac{4M}{x_m}-\frac{4q^2}{x_m^2}+\frac{q^4}{2Mx_m^3}\right)\exp\left(\frac{-q^2}{2Mx_m}\right)\right],\nonumber\\
a=-\frac{4\sqrt{2}W_mx_m\left(1-\frac{2M}{x_m}~\exp\left(-\frac{q^2}{2Mx_m}\right)~\right)}{q^2(1+W_m)\sqrt{2x_m\exp\left(-\frac{q^2}{2Mx_m}\right)+(-4M+\frac{4M}{x_m}-\frac{4q^2}{x_m^2}+\frac{q^4}{2Mx_m^3})\exp\left(-\frac{2q^2}{2Mx_m}\right)}},\nonumber\\
b_D(x_0)=-\frac{4\sqrt{2}W_mx_m\left(1-\frac{2M}{x_m}~\exp\left(-\frac{q^2}{2Mx_m}\right)~\right)\ln\left(\frac{4Mx_m}{2Mx_m-q^2}\right)}{q^2(1+W_m)\sqrt{2x_m\exp\left(-\frac{q^2}{2Mx_m}\right)+(-4M+\frac{4M}{x_m}-\frac{4q^2}{x_m^2}+\frac{q^4}{2Mx_m^3})\exp\left(-\frac{2q^2}{2Mx_m}\right)}}.
\end{align}
Similarly, the expression of regular part of the integral from eq. (\ref{eq:Ireg}) can be modified in terms of $x_m$ as
\begin{equation}
I_R(x_0)=~b_R~=~\int^1_0{g(z,x_m)dz}+O(x_0-x_m).
\end{equation}
We have evaluated the value of this integral numerically.
Now, writing
\begin{equation}\label{eq:b}
b=-\pi+b_D+b_R,
\end{equation}
we obtain the deflection angle from eqs. (\ref{eq:alpha}) , (\ref{eq:I}) , (\ref{eq:IDx0}) and (\ref{eq:b}) as
\begin{equation}
\alpha(x_0)=-a\log\left(\frac{x_0}{x_m}-1\right)+b+O(x_0-x_m).
\end{equation}
We have plotted   the deflection   angle ($\alpha(x_0)$)
 with respect to the  distance of closest approach $(x_0)$ to the
black hole in Fig 2. Also have made a comparison with the
Schwarzschild black hole  and the Reissner Nordstr\"{o}m black
hole.

\begin{figure}\label{fig:compare}
\includegraphics[width=8 cm]{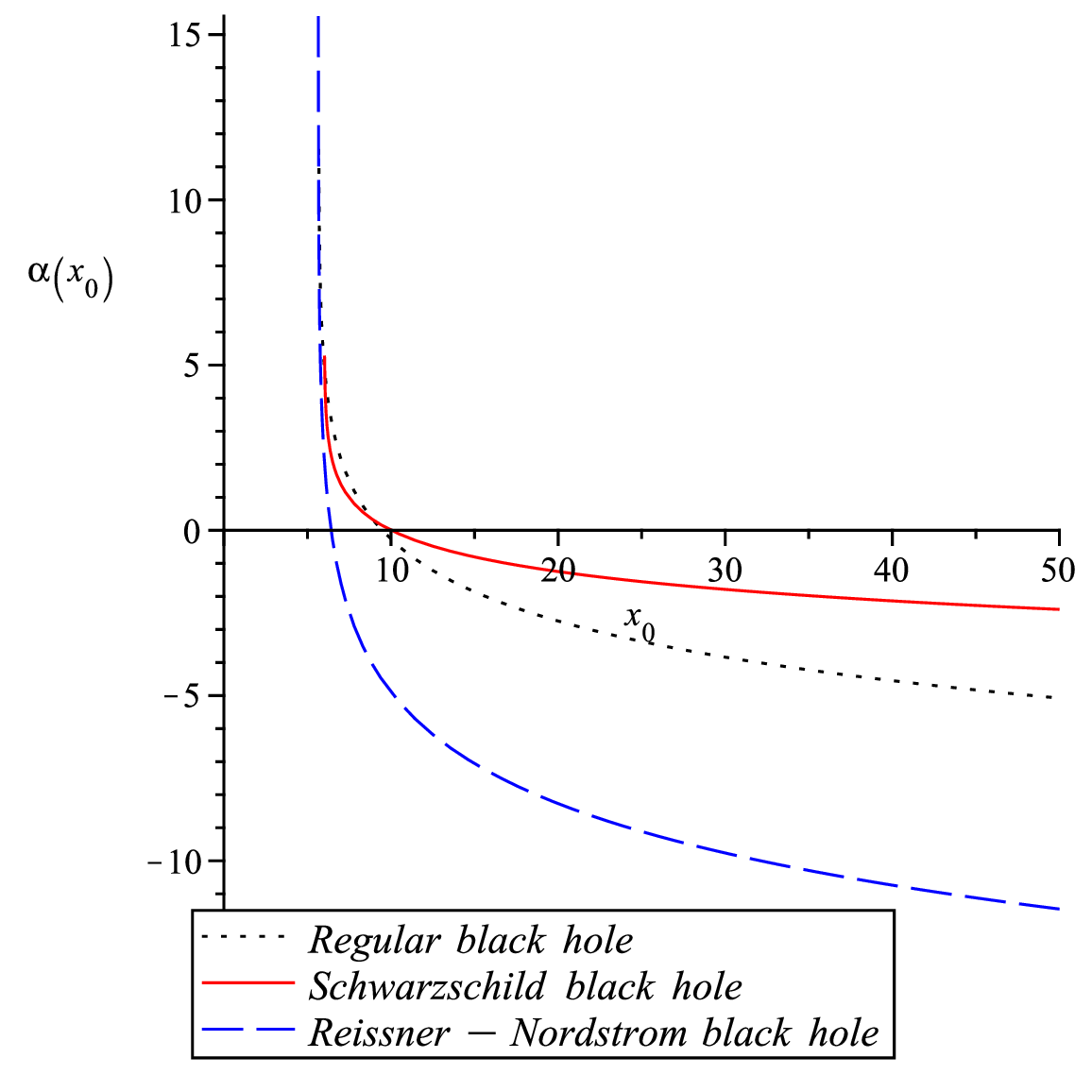}
\caption{  The angle of deflection ($\alpha(x_0)$) for  the present black hole under
study as well as the Schwarzschild black hole   and   the Reissner Nordstr\"{o}m black hole  is plotted against distance of closest approach ($x_0$)
 with the parametric values as $q=1M_\odot$ and $M=2.5M_\odot$. }
\end{figure}

 Let $D_{ol}$ be the distance between the lens and the observer and the angular separation of the image from the lens be $\theta=\frac{u}{D_{ol}}$. Recalling from eq.\ref{eq:u} we can define
\begin{align}
u_m&=\sqrt{\frac{C(x_m)}{y(x_m)}}\nonumber\\
&=\sqrt{\frac{x_m^2}{1-\frac{2M}{x_m}\exp{\left(\frac{-q^2}{2Mx_m}\right)}}}.
\end{align}
Expanding eq.\ref{eq:u} we get,
\begin{align}
u-u_m=c(x_0-x_m)^2,
\end{align}
where
\begin{equation}
c=\frac{C_m''y_m-C_my_m''}{4\sqrt{y_m^3C_m}}.
\end{equation}
Then the deflection angle can be written in terms of the angular separation $\theta$ as follows:
\begin{align}
\alpha(\theta)=-\overline{a}\log\left(\frac{\theta D_{ol}}{u_m}-1\right)+\overline{b},
\end{align}
where
\begin{align}
\overline{a}&=\frac{a}{2}\nonumber\\
&=-\frac{4\sqrt{2}W_mx_m\left(1-\frac{2M}{x_m}~\exp\left(-\frac{q^2}{2Mx_m}\right)~\right)}{2q^2(1+W_m)\sqrt{2x_m\exp\left(-\frac{q^2}{2Mx_m}\right)+(-4M+\frac{4M}{x_m}-\frac{4q^2}{x_m^2}+\frac{q^4}{2Mx_m^3})\exp\left(-\frac{2q^2}{2Mx_m}\right)}},
\end{align}
and
\begin{align}
\overline{b}&=b+\frac{a}{2}\log\left(\frac{cx_m^2}{u_m}\right)\nonumber\\
&=-\pi+b_R+\overline{a}\log\left(\frac{\frac{M^2}{x_m^2}\exp\left(\frac{-q^2}{Mx_m}\right)}{\left(1-\frac{2M}{x_m}\exp\left(\frac{-q^2}{2Mx_m}\right)\right)}\left[2x_m+\left(-4M+\frac{4M}{x_m}-\frac{4q^2}{x_m^2}+\frac{q^4}{2Mx_m^3}\right)\exp\left(\frac{-q^2}{2Mx_m}\right)\right]\right).
\end{align}
 Here we have evaluated $b_R$ numerically. In Fig 3,  we have plotted the angle $\alpha$ as a function of angular separation $\theta$ along a continuous curve along with some other characteristic features of the image, which we will discuss in details in the next section. Angles are measured in arcsecs.

\begin{figure}\label{fig:theta}
\includegraphics[width=8 cm]{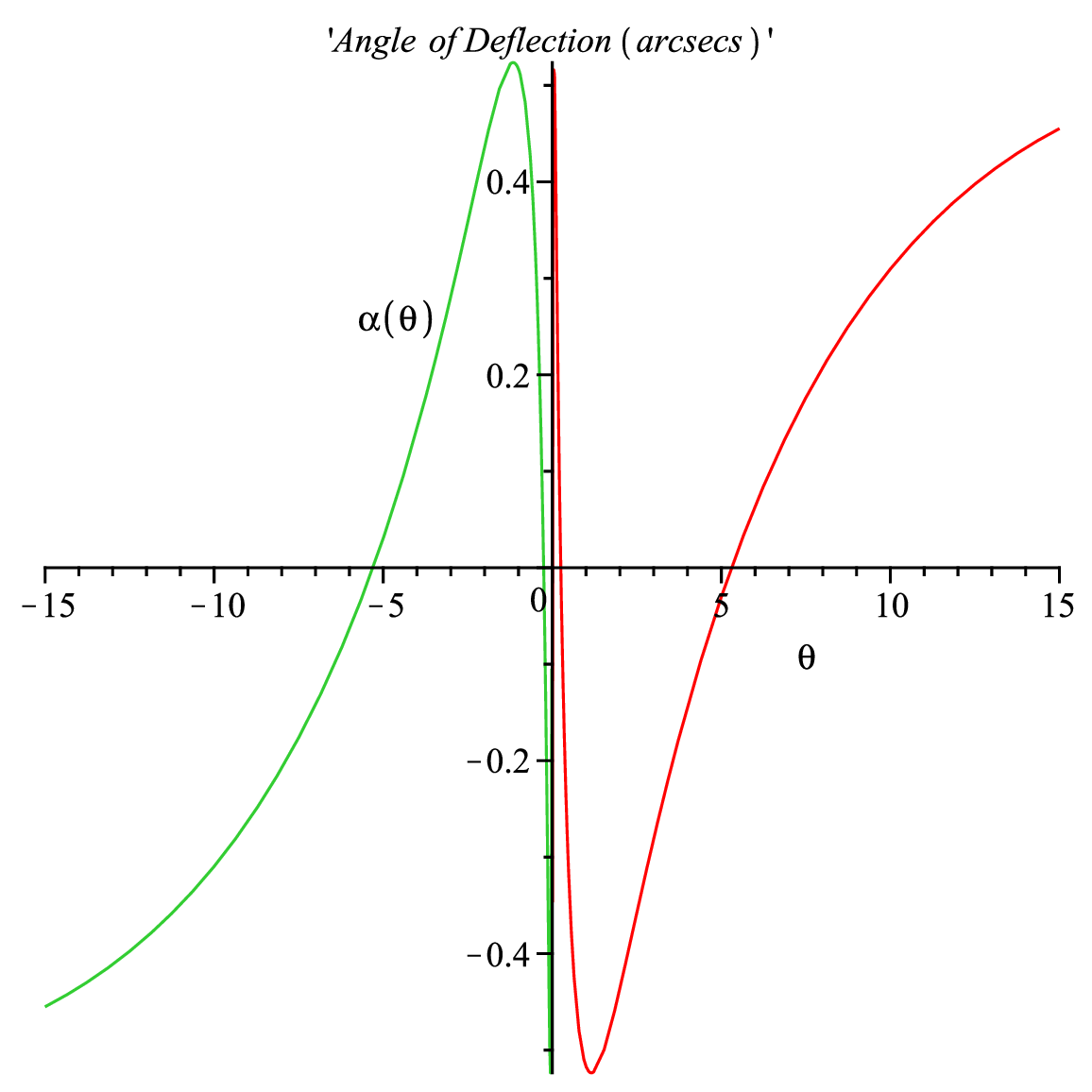}
\caption{  The angle of deflection ($\alpha(x_0)$) for  the present black hole under
study is plotted as a function of $\theta$,
 with the parametric values as $q=1M_\odot$ and $M=2.5M_\odot$. }
\end{figure}

\section{Image Analysis}
\begin{figure}\label{fig:thetaall}
\includegraphics[width=8 cm]{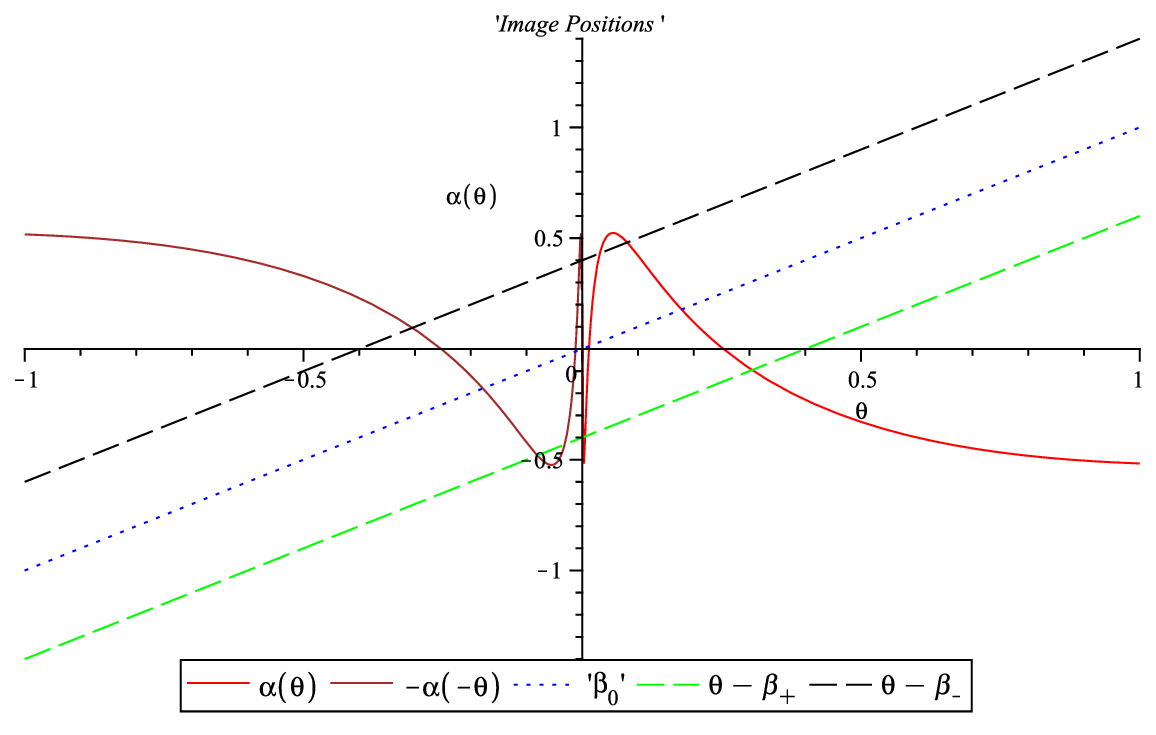}
\caption{  The angle of deflection $ y \equiv \widehat{\alpha}$ and $\theta-\beta$ as a function of $\theta$ and $\widehat{-\alpha}$ and $-\theta-\beta$ as a function of $-\theta$
 with the parametric values as $q=1M_\odot$ and $M=2.5M_\odot$. We have taken the ratio $\frac{D_{ls}}{D_os}=\frac{1}{2}$ and $D_{ol}=10 M$. The dotted straight line passing through the origin corresponds to $\beta=0$.We have plotted the lines $\theta(\theta)-\beta(\theta)$ and $-\theta(-\theta)-\beta(-\theta)$ taking $\beta=0.4 arcsecs$ .  }
\end{figure}

 In the previous section we have already defined $D_{ol}$ to be the fixed angular distance between the observer and the lens. Let $D_{os}$ be the angular distance between the observer and the source and $D_{ls}$ be the angular distance between the source and the lens. We take our reference axis to be line between observer and lens. Then as previously mentioned $\theta$ is the angular separation between the tangent to the null geodesic at the observer and the reference axis. Let $\beta$ denotes the angular separation between the tangent to the null geodesic at the true source position and the reference axis. We know the deflection angle is denoted by $\alpha$. The deflection angle is defined as positive when the lens has attractive nature and hence forms images further away from the reference axis than its true source position. And it is defined as negative when when the lens has repulsive nature and so forms images closer to the reference axis than its true source position\cite{Ellis}.\\  We can write the lens equation as \cite{Virbhadra1998}:
\begin{equation}
\sin{(\theta-\beta)}=\frac{D_{ls}}{D_{os}}\sin{\alpha(\theta)}.
\end{equation}
We know define the reduced Einstein deflection angle as,
\begin{equation}
\widehat{\alpha(\theta)}=\sin^{-1}\left(\frac{D_{ls}}{D_{os}}\sin{\alpha(\theta)}\right).
\end{equation}
Obviously we have,
\begin{equation}
\sin{(\theta)}=\frac{u}{D_{ol}},
\end{equation}
which implies,
\begin{equation}
\sin{\theta}=\frac{x_0}{D_{ol}\sqrt{1-\frac{2M}{x_0}\exp{\left(\frac{-q^2}{2Mx_0}\right)}}}.
\end{equation}
The magnification of the image is given by,
\begin{equation}
\mu=\left(\frac{\sin\beta}{\sin\theta}\frac{d\beta}{d\theta}\right)^{-1}.
\end{equation}
The tangential and radial critical curves are given by the points of singularity of the following two expressions,
\begin{equation}
\mu_t=\left(\frac{\sin\beta}{\sin\theta}\right)^{-1},
\end{equation}
and
\begin{equation}
\mu_r=\left(\frac{d\beta}{d\theta}\right)^{-1},
\end{equation}
respectively.
 We have plotted the graphs of $\widehat{\alpha}$ and $\theta-\beta$ as a function of $\theta$ and $\widehat{-\alpha}$ and $-\theta-\beta$ as a function of $-\theta$ in Fig 4. This enables us to study the images on both sides of the reference axis. We have taken the ratio $\frac{D_{ls}}{D_os}=\frac{1}{2}$ and $D_{ol}=10 M$. The dotted straight line passing through the origin corresponds to $\beta=0$. It intersects both the continuous curve $\widehat{\alpha(\theta)}$ and  $\widehat{-\alpha(-\theta)}$ in two points, these denote the positions of the tangential critical curves (TCC) or Einstein rings. Thus from Fig.[\ref{fig:thetaall}] we can find that there are two concentric Einstein rings. We have plotted the lines $\theta(\theta)-\beta(\theta)$ and $-\theta(-\theta)-\beta(-\theta)$ taking $\beta=0.4 arcsecs$ and obtained two points of intersection each with the curves  $\widehat{\alpha(\theta)}$ and $\widehat{-\alpha(-\theta)}$ respectively. These points of intersections give the location of the images on either side of the reference axis. Out of the these the two images closer to the axis are called inner image and the ones farther away are called outer image.   \\The total magnification $\mu$ is plotted in Fig 5 and the tangential $\mu_t$ and radial magnifications $\mu_r$ in Figs. 6 \&  7. The points of singularities of $\mu_t$ gives the location of tangential critical curves (TCC) and the radial critical curves (RCC). The graph shows the position of a single RCC and the magnification decreases steeply as $\theta$ is further increased.

\begin{figure}\label{fig:mu}
\includegraphics[width=10 cm]{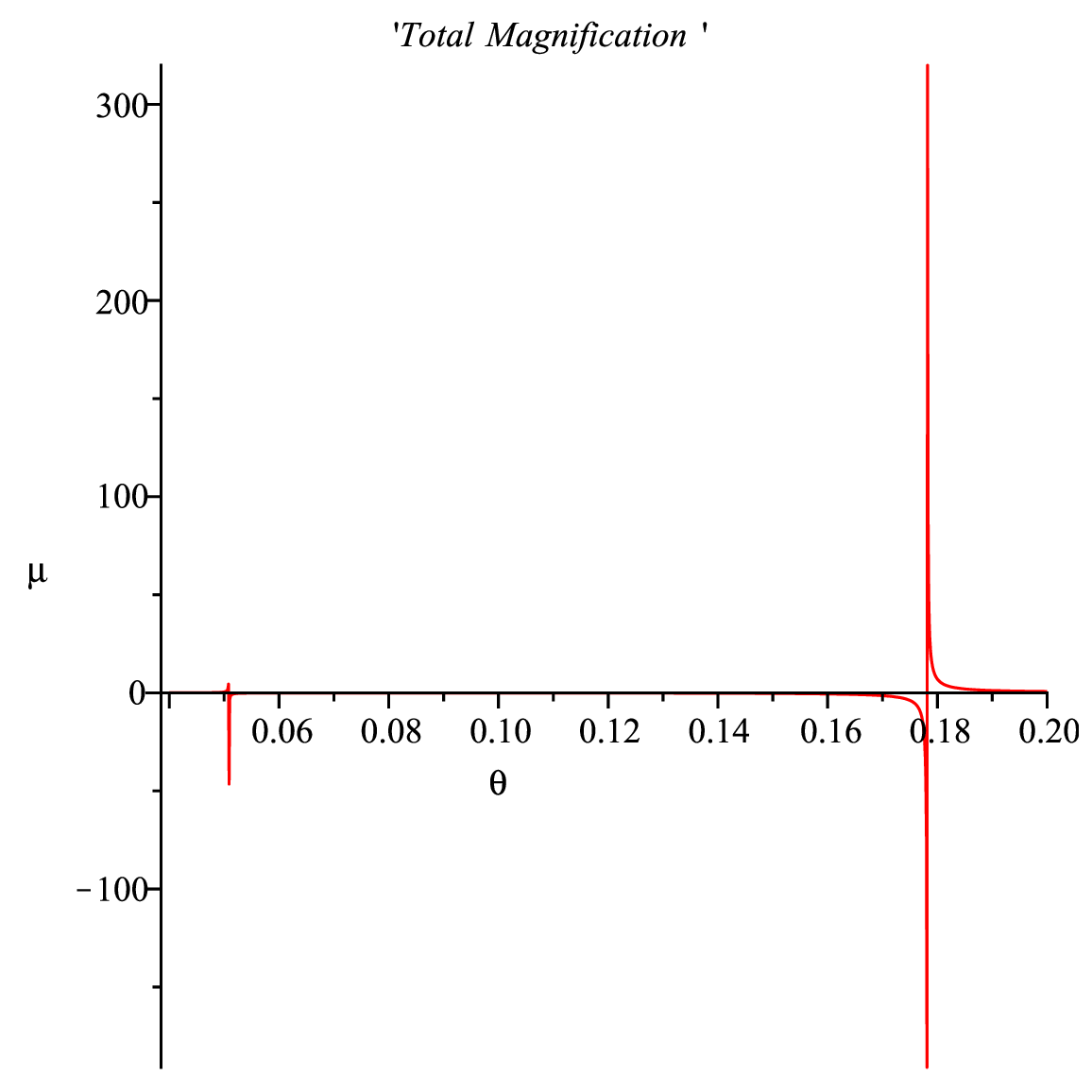}
\caption{  The total magnification $\mu$ is plotted
 with the parametric values as $q=1M_\odot$ and $M=2.5M_\odot$. We have taken $D_{ol}=10 M$, $\beta=0.4 arcsecs$ .  }
\end{figure}

\begin{figure}\label{fig:murt}
\includegraphics[width=10 cm]{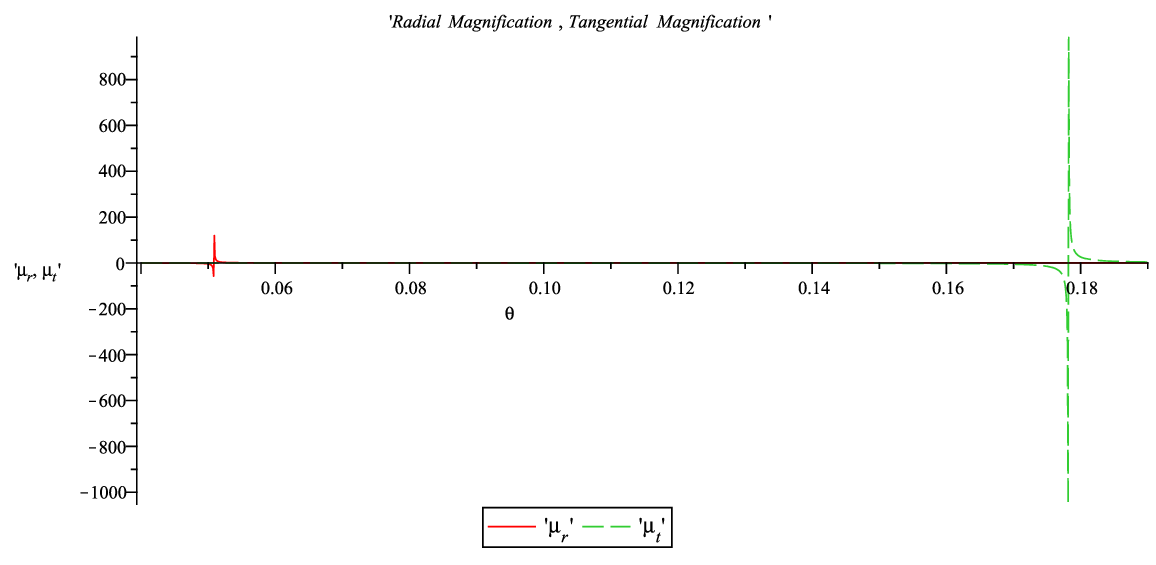}
\caption{  The tangential $\mu_t$ and radial magnifications $\mu_r$
 with the parametric values as $q=1M_\odot$ and $M=2.5M_\odot$. We have taken $D_{ol}=10 M$, $\beta=0.4 arcsecs$ .The points of singularities of $\mu_t$ gives the location of tangential critical curves (TCC) and the radial critical curves (RCC). The graph shows the position of a single RCC and the magnification decreases steeply as $\theta$ is further increased.  }
\end{figure}

\begin{figure}\label{fig:murt2}
\includegraphics[width=10 cm]{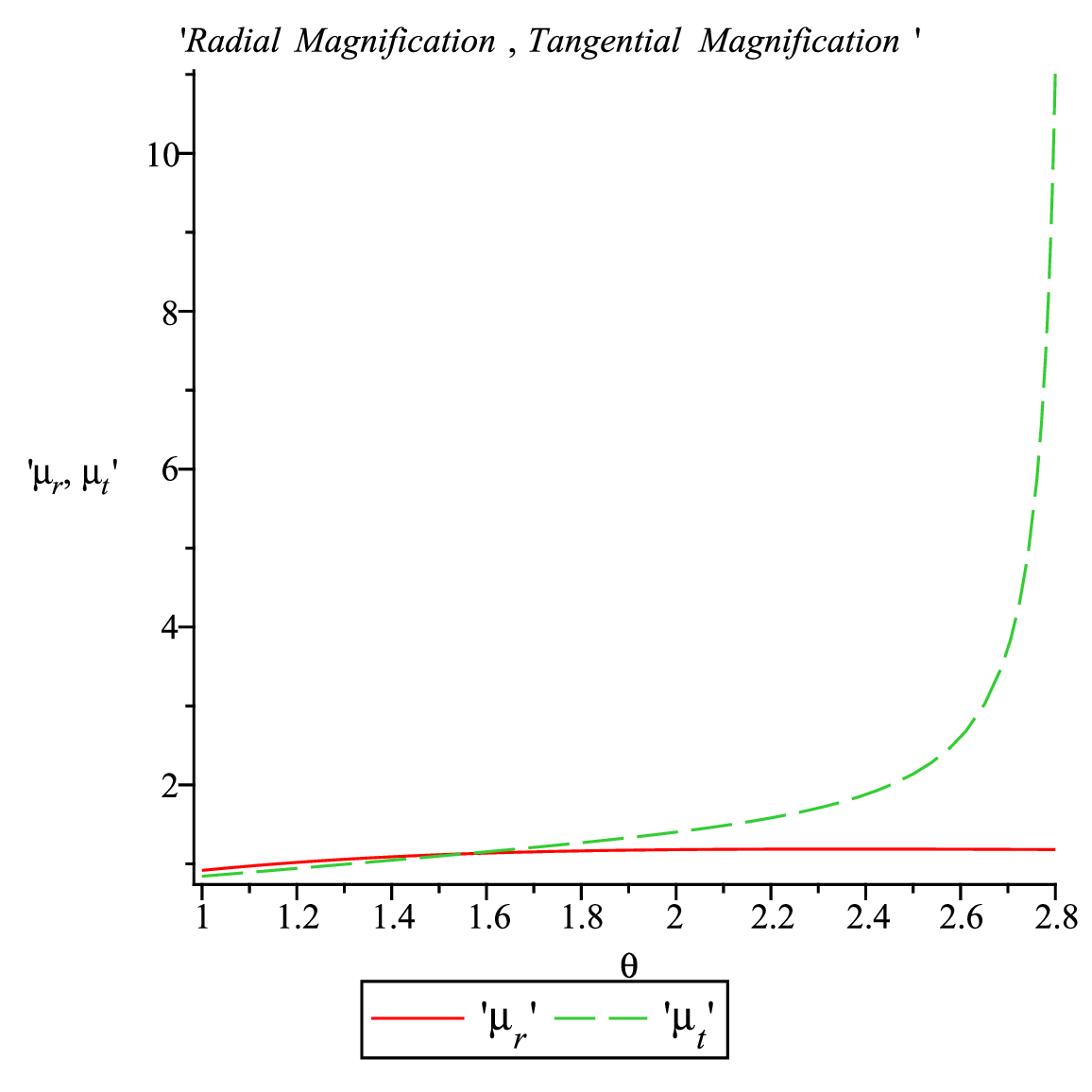}
\caption{  The tangential $\mu_t$ and radial magnifications $\mu_r$
 with the parametric values as $q=1M_\odot$ and $M=2.5M_\odot$. We have taken $D_{ol}=10 M$, $\beta=0.4 arcsecs$ .The points of singularities of $\mu_t$ gives the location of tangential critical curves (TCC) and the radial critical curves (RCC). This graph shows the absence of any RCC but TCC are present.  }
\end{figure}

\section{Concluding Remarks}

 In this paper, we have studied the strong lensing effect
after considering spherically symmetric regular black hole.
Graphically (see Fig 1)  we have found the radius of photon sphere
as $x_m=5.6559907$ for $M=2.5M_\odot$ and $q=1 M_\odot$. In our study the
analytical expression for deflection angle due to the strong
lensing effect has been found and then graphically represented in
Fig 2.  From this figure we can conclude that the deflection angle
$\alpha(x_0)$   is highly positive when the value of the distance
of closest approach $(x_0)$ is close to the value of photon sphere
($x_m$). As $x_0$ equals
the photon sphere the path of light loops infinitely around the
black hole, but as $x_0$ increases beyond $x_m$ the angle of
deflection decreases rapidly as the light undergoes gradually
lesser bending, until at a particular value of $x_0$ there is no
deflection of light at all; this corresponds graphically to the
point $x_0=9.46$ for the black hole under study and to the point
$x_0=10$ for the Schwarzschild black hole  and and to the point $x_0=6.42$ for the Reissner Nordstr\"{o}m black hole. For $x_0$ greater than
these particular values the angle of deflection is found to be
negative, which can be interpreted as there is a repulsion of
light by the corresponding black hole.  A possible explanation can be in terms of refractive index of the material medium. We can assume that there exists two concentric shells around the photon sphere; the inner shell  consisting of a denser material whose refractive index is higher than exterior spacetime and the outer shell is made up of a rarer material whose refractive index is lower than exterior spacetime. This leads to overall positive deflection as light travels from a rarer to denser medium near the photon sphere, reinforced further by the immense gravity of the black hole; whereas away from the photon sphere, in the outer shell, light travels from a denser to a rarer medium and hence gets repulsed. However this idea requires further effort with detailed calculations to verify it. Some future work can be done on this direction.

\begin{acknowledgements}
FR  would like to thank the authorities of the Inter-University Centre for Astronomy
and Astrophysics, Pune, India for providing the research facilities. FR and SM are  also
thankful to DST-SERB and CSIR, Govt. of India for financial support.
\end{acknowledgements}

\end{document}